\documentclass[conference]{IEEEtran}
\IEEEoverridecommandlockouts
\usepackage{cite}
\usepackage{amsmath,amssymb,amsfonts}
\usepackage{algorithmic}
\usepackage{graphicx}
\usepackage{textcomp}
\usepackage{xcolor}
\usepackage{multirow}
\usepackage{pgfplots}
\usepackage{xurl}

\usetikzlibrary{patterns}

\pgfplotsset{compat=1.18}
\usepgfplotslibrary{groupplots}
\usepgfplotslibrary{fillbetween}

\def\BibTeX{{\rm B\kern-.05em{\sc i\kern-.025em b}\kern-.08em
    T\kern-.1667em\lower.7ex\hbox{E}\kern-.125emX}}
\begin{document}

\title{Active Inference for Adaptive Traffic Signal Control in Noisy Nonstationary IoT Environments\\
\thanks{
\IEEEauthorrefmark{2}Authors contributed equally to this work.
}
}

\author{
\IEEEauthorblockN{Dénes Toth\IEEEauthorrefmark{2}, George Ambroladze\IEEEauthorrefmark{2}, Edwin Sundberg, Ali Beikmohammadi, and Alfreds Lapkovskis}
\IEEEauthorblockA{\textit{Department of Computer Systems and Sciences}, \textit{Stockholm University,} Stockholm 164 25, Sweden}
\IEEEauthorblockA{\texttt{\{toth.denes1419, giorgi.ambroladze\}@gmail.com}}
\IEEEauthorblockA{\texttt{\{edwinsu, beikmohammadi, alfreds.lapkovskis\}@dsv.su.se}}
}

\maketitle

\begin{abstract}
Urban traffic signal control at IoT-instrumented intersections must remain effective under sensor occlusion, weather attenuation, and nonstationary demand. Conventional controllers degrade under these conditions, and learned policies remain difficult to audit. 
To address these challenges,
we propose an 
active inference controller for a
four-arm signalized intersection that dynamically selects phases by minimizing expected free energy (EFE) over Gaussian beliefs about per-direction congestion levels,
yielding a fully traceable decision pipeline. 
We benchmark the controller
in a SUMO traffic simulator against a rule-based heuristic and a deep Q-network (DQN) across four scenarios that progressively increase noise and nonstationarity, spanning sensor occlusion, adverse weather, and stochastic accidents. Across 100 independent random evaluations per scenario, active inference attains the lowest idle times and CO2 emissions in the noisiest scenarios 
(56{,}977\,s and 29.12\,kg vs. 71{,}741\,s and 30.56\,kg for DQN).
These gains come at a modest cost in bus priority service rate and phase switch frequency.
\end{abstract}


\section{Introduction}
Urban traffic congestion and heavy traffic negatively impact air pollution levels and, as a consequence, human health 
\cite{schrank2019urban,levy2010evaluation,raysoni2009health}. Furthermore, congestion and heavy traffic affect economic activity and the likelihood of road accidents \cite{sweet2011does,wang2009impact}. One established countermeasure is adaptive traffic signal control \cite{papageorgiou2003review}, whose primary objective is to improve mobility, reduce congestion, and enhance network efficiency under traffic dynamics \cite{wang2023critical}. Smart traffic-light management is a promising approach for reducing waiting times and improving fuel economy on congested networks \cite{patidar2021optimizing}.

Approaches to traffic control range from rule-based methods \cite{dion2002rule,feng2015real,guo2024scalable} to predictive model-based techniques \cite{wei2024research}, which anticipate future states rather than solely react to current observations. More recently, deep reinforcement learning has become the dominant paradigm \cite{rasheed2020deep,wei2018intellilight}. Both reinforcement learning and model predictive control achieve adaptivity by updating actions based on current observations or predicted future states \cite{ye2019survey}. Rule-based methods are straightforward to implement and interpret because they rely on predefined signal behavior, yet lack the flexibility to handle unpredictable conditions \cite{gheorghe2025revolutionizing}. In contrast, deep reinforcement learning achieves strong performance but demands large training datasets, and its black-box nature raises concerns about interpretability \cite{thadikamalla2025reinforcement}. In practice, however, most real-world intersections remain governed by fixed-time schedules or rule-based logic triggered by IoT sensors \cite{wei2019survey}. No universally accepted method for designing intelligent traffic lights has emerged, as the right choice depends on traffic characteristics and road network topology \cite{patidar2021optimizing}. A further challenge is that traffic measurements are often incomplete or noisy, requiring controllers to act on imperfect information \cite{aslani2017adaptive}.

Precisely for such conditions, active inference (AIF) offers a principled alternative by explicitly accounting for uncertainty and partial observability \cite{friston2006free,friston2021sophisticated}. It frames perception and action as a single process of minimizing variational free energy, that is, 
a tractable bound on how much its internal world model fails to account for incoming observations.
AIF has been successfully applied to control problems outside of traffic, including robotic systems \cite{lanillos2021active},
yet to the best of our knowledge, it has not been explored in traffic signal control.

Intelligent controllers rely on sensor networks, camera systems, or simulated observations to estimate queue lengths, vehicle arrivals, and flow patterns \cite{deshpande2023cyber}. Because traffic control objectives such as minimizing delay, reducing emissions, and prioritizing public transport often conflict, they must balance multiple performance criteria simultaneously \cite{uribe2025evaluating}. Performance is typically reported on standard metrics such as average vehicle delay, queue length, intersection throughput, travel time, and number of stops per vehicle \cite{kaparias2011key}, which support systematic comparison between intelligent controllers and baseline methods such as fixed-time, actuated, or rule-based systems \cite{dion2004comparison}. In real deployments, these controllers operate at the edge of IoT-instrumented intersections, where vehicle counts are inferred from camera, LiDAR, or loop detectors over lossy and bandwidth-constrained links \cite{deshpande2023cyber,anand2020region}. The resulting sensor channel is noisy and partial: occlusion by heavy vehicles, adverse weather, and intermittent connectivity all degrade observation quality at exactly the moments when control quality matters most.

We therefore propose an AIF traffic signal controller and evaluate its behavior under noise and nonstationarity introduced by real-world IoT sensing conditions, comparing it against rule-based and deep reinforcement learning baselines across multiple performance dimensions. Our contribution is threefold:
\begin{itemize}
    \item We design a simulation environment that models noisy, nonstationary IoT sensing conditions at a signalized intersection, including sensor occlusion, adverse weather attenuation, and stochastic traffic accidents.
    \item We develop an AIF controller that dynamically selects phases by minimizing EFE over Gaussian beliefs about congestion, emissions, and public transport presence.
    \item We evaluate this controller against a rule-based heuristic and DQN \cite{mnih2015human} across four scenarios of increasing environmental disturbance, identifying and discussing the tradeoffs inherent to EFE-based action selection.
\end{itemize}

The remainder of this paper is organized as follows. Section \ref{sec:proposed_method} presents the proposed controller, Section \ref{sec:experiment} describes the experimental setup, Section \ref{sec:results} reports results, Section \ref{sec:discussion} discusses findings, and
Section \ref{sec:conclusion} concludes the paper.

\section{Proposed Method} \label{sec:proposed_method}
At each discrete time step $t$, an AIF agent perceives the environment through observations $o_t$, maintains approximate posterior beliefs $q(s_t)\approx p(s_t\mid o_t)$ over hidden states $s_t$, and acts by sampling actions $a_t\sim \pi$, where the policy $\pi$ is a sequence of actions that minimizes EFE $\mathcal{G}(\pi)$, defined as
\begin{equation}
    \mathcal{G}(\pi)\triangleq-[\mathrm{PV}(\pi)+\lambda\,\mathrm{EV}(\pi)],
\end{equation}
where $\lambda\in[0,1]$ is a constant that weights the importance of $\mathrm{EV}(\cdot)$. Here  $\mathrm{PV}(\cdot)$ is a \emph{pragmatic value} that rewards observations matching our preferences, defined as
\begin{equation}
    \mathrm{PV}(\pi)\triangleq \sum_{\tau=t+1}^T \gamma^{\tau-t-1}\underbrace{\mathbb{E}_{q(o_\tau \mid \pi)}[\log P(o_\tau)]}_{\triangleq pv(\pi, \tau)},
    \label{eq:pv}
\end{equation}
where $\gamma\in(0,1)$ is a discount factor that downweights contributions from future time steps, and $P(\cdot)$ encodes preferences over observations, and $T\rightarrow\infty$ is a time horizon;
and $\mathrm{EV}(\cdot)$ is an \emph{epistemic value} that rewards observations that reduce belief uncertainties, and is defined as
\begin{equation}
    \mathrm{EV}(\pi)\triangleq\hspace{-0.5em}\sum_{\tau=t+1}^T\hspace{-0.2em} \gamma^{\tau-t-1} \mathbb{E}_{q(o_\tau\mid\pi)}[ D_{KL}(q(s_\tau{\mid} o_\tau,\pi)\lVert q(s_\tau\mid\pi))],
    \label{eq:ev}
\end{equation}
where $D_{KL}(\cdot)$ is Kullback-Leibler divergence \cite{friston2006free,friston2015active,friston2017active}.

We instantiate this action-perception framework as a single-agent controller for a four-arm signalized intersection. Traffic is aggregated into a North-South (NS) and East-West (EW) approach pair, and the controller selects one of two phases at each decision step, $a_t \in \{\text{NS green}, \text{EW green}\}$. Per direction, hidden states $s_t$ are discretized into six congestion levels (1: very-low through 6: jam) with bin boundaries derived from empirical traffic distributions. The observations $o_t$ are defined as
vectors $\mathbf{o}_t = (n_t, c_t, b_t) \in \mathbb{R}^3$ with the noisy vehicle count $n_t$, the cumulative CO\textsubscript{2} emission $c_t$, and the bus count $b_t$. The observation model is a state-conditional multivariate Gaussian 
\begin{equation}
    (\mathbf{o}_t \mid s_t=s_i) \sim \mathcal{N}(\boldsymbol{\mu}_i, \boldsymbol{\Sigma}_i),
    \label{eq:observation_model}
\end{equation}
whose means $\boldsymbol{\mu}_i$ and covariances $\boldsymbol{\Sigma}_i$ are fitted to per-state simulator samples (means $\boldsymbol{\mu}_i$ tabulated in Table~\ref{tab:state_means}). Two manually specified action-conditioned transition matrices, $\mathbf{T}_{\text{red}}$ and $\mathbf{T}_{\text{green}}$ (Figs.~\ref{fig:t_red} and \ref{fig:t_green}), encode the intuition that a green phase concentrates row mass on equal-or-lower congestion levels and a red phase on equal-or-higher ones; the same kernel is reused across lanes despite asymmetric demand.

We encode preferences over observations as $P(o_t)=\mathcal{N}(o_t;\boldsymbol{\mu}^*,\boldsymbol{\Sigma}^*)$. Since the observation model is per-state Gaussian (Eq.~\eqref{eq:observation_model}), the expectation over observations (Eq.~\eqref{eq:pv}) can be marginalized analytically over hidden states:
\begin{equation}
    \mathbb{E}_{q(o_\tau\mid \pi)}[\log P(o_\tau)]\hspace{-0.2em}=\hspace{-0.5em}\sum_i \hspace{-0.2em}q(s_\tau{=}s_i{\mid}\pi)\underbrace{\hspace{-0.1em}\mathbb{E}_{o_\tau\sim\mathcal{N}(\boldsymbol{\mu}_i,\boldsymbol{\Sigma}_i)}[\log \hspace{-0.15em}P(o_\tau)]}_{\triangleq \log P(s_i)},
\end{equation}

therefore, $pv(\pi,\tau)\equiv \mathbb{E}_{q(s_\tau\mid\pi)}[\log P(s_\tau)]$.
This allows us to score preference matching of observations by inferred hidden states, as
\begin{equation}
\begin{aligned}
\log P(s_i) = -\tfrac{1}{2}\bigl(&\,d\log 2\pi + \log|\boldsymbol{\Sigma}^*| + \mathrm{Tr}\bigl((\boldsymbol{\Sigma}^*)^{-1}\boldsymbol{\Sigma}_i\bigr) \\
&+ (\boldsymbol{\mu}^* - \boldsymbol{\mu}_i)^{\!\top} (\boldsymbol{\Sigma}^*)^{-1}(\boldsymbol{\mu}^* - \boldsymbol{\mu}_i)\bigr),
\end{aligned}
\label{eq:pref_score}
\end{equation}
where $d=3$ is the observation-space dimensionality. 
Additionally,  we weight $pv$ terms by expected proportions $\tilde{s}_d$ of NS/EW traffic load, so that the busier direction is prioritized, i.e., redefining $pv$ as $pv(\pi,\tau)=\sum_{d\in\{\mathrm{NS},\mathrm{EW}\}}\tilde{s}_{d}\,pv_d(\pi,\tau)$.

Our AIF agent samples actions $a_t$ greedily, mirroring the identical sampling strategy of our DQN baseline used in evaluation.
Unlike a DQN whose policy is encoded in the opaque weights of a function approximator, every stage of the AIF controller is explicit and traceable: the posterior belief, the per-action predicted belief, the pragmatic and epistemic components of EFE, and the resulting action posterior can each be inspected at every decision step. An operator auditing a phase choice can therefore read off which component of the EFE dominated for the selected action.

\begin{figure}[!tb]
\centering
\footnotesize
$\mathbf{T}_{\text{red}} = \begin{bmatrix}
0.40 & 0.55 & 0.04 & 0.01 & 0.00 & 0.00 \\
0.05 & 0.30 & 0.55 & 0.09 & 0.01 & 0.00 \\
0.00 & 0.00 & 0.20 & 0.70 & 0.08 & 0.02 \\
0.00 & 0.00 & 0.00 & 0.20 & 0.60 & 0.20 \\
0.00 & 0.00 & 0.00 & 0.00 & 0.70 & 0.30 \\
0.00 & 0.00 & 0.00 & 0.00 & 0.10 & 0.90
\end{bmatrix}$
\caption{Action-conditioned transition matrix $\mathbf{T}_{\text{red}}$ over the six congestion levels (1: very-low, 6: jam); entry $[i,j]$ is the probability of transitioning from state $i$ to state $j$ while the direction is red.}
\label{fig:t_red}
\end{figure}

\begin{figure}[!tb]
\centering
\footnotesize
$\mathbf{T}_{\text{green}} = \begin{bmatrix}
0.80 & 0.18 & 0.02 & 0.00 & 0.00 & 0.00 \\
0.15 & 0.60 & 0.22 & 0.03 & 0.00 & 0.00 \\
0.08 & 0.32 & 0.44 & 0.14 & 0.02 & 0.00 \\
0.02 & 0.08 & 0.33 & 0.42 & 0.13 & 0.02 \\
0.01 & 0.03 & 0.33 & 0.43 & 0.15 & 0.05 \\
0.05 & 0.10 & 0.25 & 0.30 & 0.20 & 0.10
\end{bmatrix}$
\caption{Action-conditioned transition matrix $\mathbf{T}_{\text{green}}$ over the six congestion levels; entry $[i,j]$ is the probability of transitioning from state $i$ to state $j$ while the direction is green.}
\label{fig:t_green}
\end{figure}

\begin{table}[!tb]
\centering
\caption{Per-state observation means $\boldsymbol{\mu}_i = (n, c, b)$ in (count, CO\textsubscript{2} mg/s, bus count) coordinates, used together with the per-state covariances $\boldsymbol{\Sigma}_i$ (fitted to simulator samples; omitted for space) in the Gaussian likelihood $\mathbf{o} \mid s_i \sim \mathcal{N}(\boldsymbol{\mu}_i, \boldsymbol{\Sigma}_i)$.}
\label{tab:state_means}
\footnotesize
\begin{tabular}{@{}llrrr@{}}
\hline
$i$ & state & $n$ & $c$ & $b$ \\
\hline
1 & very-low   &  2.98 &  7.85 & 0.52 \\
2 & low        &  9.13 & 22.39 & 1.17 \\
3 & medium     & 15.16 & 33.17 & 1.94 \\
4 & high       & 21.25 & 49.07 & 2.63 \\
5 & extra-high & 28.28 & 60.12 & 2.78 \\
6 & jam        & 39.05 & 73.24 & 2.59 \\
\hline
\end{tabular}
\end{table}

\section{Experimental Design} \label{sec:experiment}
\begin{figure}[!htb]
\centering
\includegraphics[width=0.6\columnwidth]{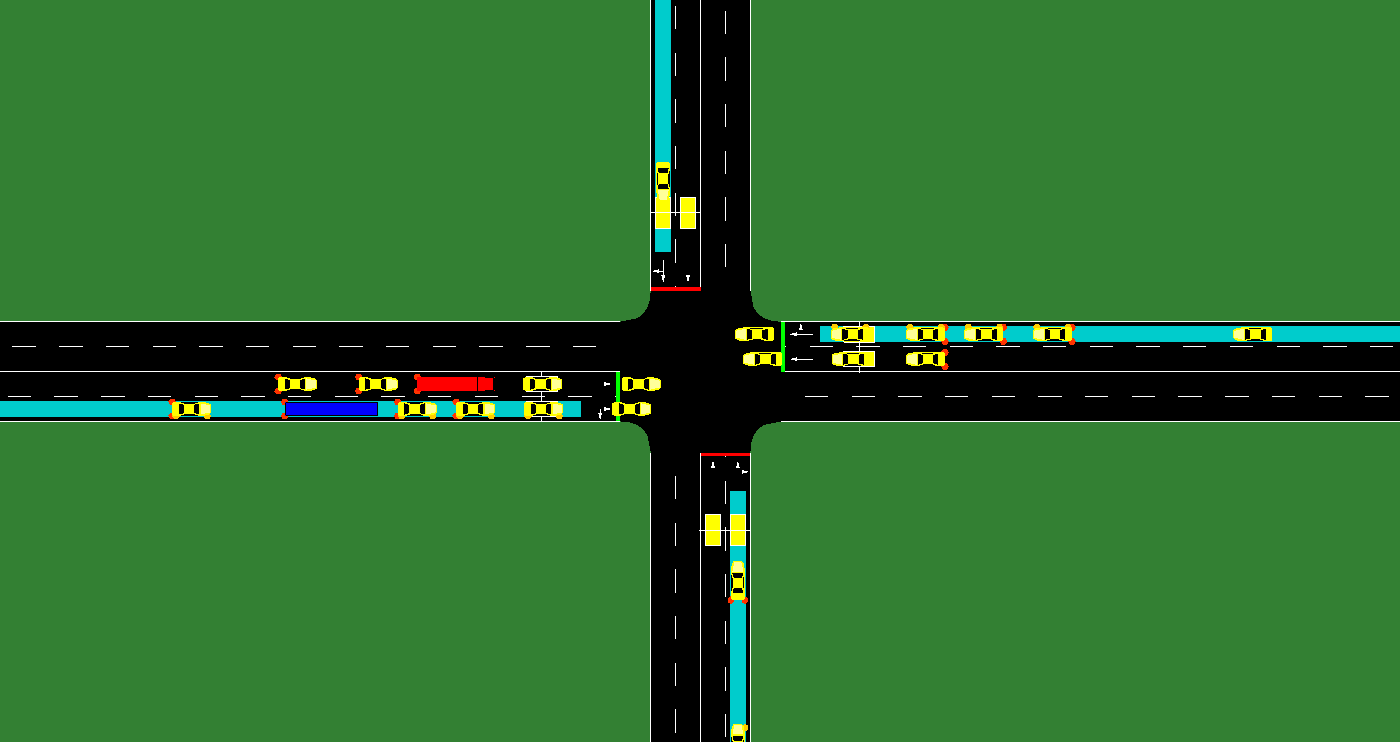}
\caption{SUMO render of the four-arm signalized intersection used throughout the evaluation. The main road runs east--west and the secondary road north--south, with queues forming on all four approaches during a peak-demand step. Yellow vehicles are passenger cars, the red vehicle is a truck, and the blue vehicle is a bus. The asymmetric east--west loading is characteristic of the demand profile and underlies the bus priority and emissions trade-offs reported below.}
\label{fig:sumo_intersection}
\end{figure}

We conducted experiments using Simulation of Urban MObility (SUMO) \cite{lopez2018microscopic}, a well-established open-source microscopic traffic simulator, on a single four-arm intersection with incoming and outgoing lanes in each direction (Fig.~\ref{fig:sumo_intersection}). Traffic is heterogeneous, consisting of passenger cars, buses, and trucks, in accordance with the modeling assumptions of \cite{jabari2012stochastic}. Buses carry public priority and produce moderate emissions ($\approx$11{,}519\,mg/veh), trucks produce the highest emissions ($\approx$12{,}922\,mg/veh), and passenger cars the lowest ($\approx$2{,}990\,mg/veh), and both heavy classes contribute to perceptual occlusion. Demand follows a full-day nonstationary profile (low early hours, morning peak, midday plateau, evening peak, late-night decay), capturing the time-of-day variation that drives most urban traffic dynamics \cite{patidar2021optimizing}; a main road and secondary road share that profile, with the secondary road carrying lower flow and fewer heavy vehicles, creating asymmetric demand across the intersection.

Real intersection sensors operate far from perfect detection: a representative LiDAR-based vehicle-detection pipeline for advanced traffic management reports only 64.5\% average detection accuracy on the KITTI benchmark \cite{anand2020region}, and roadside camera/LiDAR deployments suffer compounding losses under occlusion and adverse weather. The perception stack is therefore modeled as a lossy IoT sensor channel rather than a direct read-out of true counts: observations are generated by a two-stage stochastic process. Let $N^{\text{true}}_d$ be the true vehicle count in direction $d \in \{\text{NS}, \text{EW}\}$ and $K_d$ the number of heavy vehicles. A per-vehicle detection probability
\begin{equation}
p_d = \nu^{w}\,\bigl(1 - \alpha\,\max(1, N^{\text{true}}_d - 1)\bigr)
\label{eq:pd}
\end{equation}
combines a weather attenuation 
$\nu=0.8$
active when the binary weather indicator $w = 1$ and a density-dependent term scaled by $\alpha=0.005$, and a binomial draw $\tilde{N}_d \sim \mathrm{Binomial}(N^{\text{true}}_d, p_d)$ determines how many true vehicles are detected. Under sufficient congestion, $N^{\text{true}}_d \geq 3K_d$, an additional occlusion term is subtracted,
\begin{equation}
\begin{gathered}
O_d = \sum_{i=1}^{K_d} o_i,\quad o_i \sim \mathcal{D}_{\text{occ}}, \\
\mathcal{D}_{\text{occ}}\bigl(\{0,1,2,3\}\bigr) = (0.50, 0.30, 0.15, 0.05),
\end{gathered}
\label{eq:occlusion}
\end{equation}
yielding the final observed count $N^{\text{obs}}_d = \max\bigl(0, \tilde{N}_d - O_d\bigr)$. Stochastic traffic accidents are injected at a random time step by selecting the East-West vehicle nearest to the intersection, freezing it for a fixed duration to simulate a blockage, and then removing it.

We compare AIF against two baselines.
The \textbf{rule-based baseline} evaluates every 10\,s, computing per-direction priority scores $S_{\text{NS}} = O_{\text{NS}} + \beta_{\text{bus}} \cdot B_{\text{NS}}$ and $S_{\text{EW}} = O_{\text{EW}} + \beta_{\text{bus}} \cdot B_{\text{EW}} + \beta_{\text{main}}$, where $O_d$ and $B_d$ are the observed vehicle and bus counts in direction $d \in \{\text{NS}, \text{EW}\}$, with a fixed bus bonus $\beta_{\text{bus}} = 5$ and a main-road bias $\beta_{\text{main}} = 2$ favouring the EW direction, and switches phase whenever the opposing score is larger, subject to minimum and maximum hold times of $k_{\text{min}} = 2$ and $k_{\text{max}} = 6$ evaluation steps.

The \textbf{DQN baseline} 
has
a feedforward network of two hidden layers of 128 ReLU units each, mapping a six-dimensional state (observed NS/EW counts, cumulative NS/EW CO\textsubscript{2}, NS/EW bus counts) to two Q-values, with a per-decision reward $-(\mathrm{idle}_{\mathrm{NS}}+\mathrm{idle}_{\mathrm{EW}}) - 10^{-4}(\mathrm{CO}_{2,\mathrm{NS}}+\mathrm{CO}_{2,\mathrm{EW}}) - 0.5(\mathrm{bus}_{\mathrm{NS}}+\mathrm{bus}_{\mathrm{EW}})$ that penalizes idle, emissions, and bus load uniformly. It is trained with experience replay (buffer $50{,}000$, batch $64$), a target network refreshed every $360$ steps, discount $\gamma = 0.99$, learning rate $10^{-3}$, and an $\epsilon$-greedy policy decaying linearly from $1.0$ to $0.05$ over the first $10\%$ of training episodes. Training is run from scratch per scenario for $1{,}000$ episodes in Scenarios~1 and~3 (scenarios are defined later) and $500$ episodes in Scenarios~2 and~4 (until convergence).
Evaluation uses $\epsilon = 0$. 

The \textbf{AIF controller} updates its beliefs over hidden traffic states during evaluation but does not learn its model parameters online, so it has no analogous pretraining phase. It uses the same $\gamma=0.99$ as DQN, and $\lambda=0.5$ to reduce exploration.

Each scenario was evaluated for 3600 seconds of in-simulation time, with metrics collected at one-second intervals. Performance is reported on four metrics: mean vehicle idle time, total CO\textsubscript{2} emissions, bus priority service rate, and phase switch frequency, jointly capturing efficiency, environmental impact, prioritization of public transport, and policy stability.

Robustness is probed across four scenarios that progressively activate three independently toggleable IoT failure modes: sensor occlusion, adverse-weather attenuation, and stochastic infrastructure disruption. Scenario~1 disables all disturbances; Scenario~2 enables occlusion only; Scenario~3 additionally enables adverse weather; Scenario~4 enables all three, yielding a highly dynamic and partially observable environment. Each (scenario, controller) cell of the evaluation was replicated across 100 independent SUMO seeds. The per-scenario rankings reported below are the across-seed mean (with standard deviation and coefficient of variation in Table~\ref{tab:results_full}).

For full reproducibility, we publish our code at the GitHub repository.\footnote{\url{https://github.com/GMAN226/Python-Code-Active-Inference-for-Intelligent-Traffic-Light-Control}}

\section{Results} \label{sec:results}
We report performance across the four scenarios on idle time, CO\textsubscript{2} emissions, bus priority service rate, and phase switch frequency. Aggregate results are summarized in Table~\ref{tab:results_full} and Figs.~\ref{fig:scenario_idle} and \ref{fig:scenario_co2}.

\pgfplotsset{
    scenbars/.style={ybar, bar width=5pt, enlarge x limits=0.18,
        symbolic x coords={S1, S2, S3, S4}, xtick=data,
        width=\columnwidth, height=4cm,
        area legend,
        label style={font=\footnotesize},
        tick label style={font=\footnotesize},
        legend cell align=left,
        legend style={font=\footnotesize, draw=none, fill=none,
            at={(0.02,0.98)}, anchor=north west},
        legend image code/.code={
            \draw[#1] (0cm,-0.08cm) rectangle (0.28cm,0.12cm);
        },
        cycle list={
            {fill=blue, draw=blue, postaction={pattern=north east lines, pattern color=white}},
            {fill=green!55!black, draw=green!55!black, postaction={pattern=horizontal lines, pattern color=white}},
            {fill=orange, draw=orange, postaction={pattern=crosshatch, pattern color=white}},
        },
    },
}

\begin{figure}[!tb]
\centering
\begin{tikzpicture}
\begin{axis}[scenbars, width=1\columnwidth,  ylabel={Avg.\ idle time (s)}, ymin=0]
\addplot+[error bars/.cd, y dir=both, y explicit,
          error bar style={line width=0.4pt, black}]
    coordinates {(S1,38.93) +- (0,1.65) (S2,39.56) +- (0,2.02) (S3,39.89) +- (0,2.19) (S4,198.48) +- (0,61.54)};
\addplot+[error bars/.cd, y dir=both, y explicit,
          error bar style={line width=0.4pt, black}]
    coordinates {(S1,19.51) +- (0,3.05) (S2,23.07) +- (0,2.03) (S3,14.59) +- (0,0.87) (S4,197.29) +- (0,55.09)};
\addplot+[error bars/.cd, y dir=both, y explicit,
          error bar style={line width=0.4pt, black}]
    coordinates {(S1,20.10) +- (0,1.31) (S2,20.21) +- (0,1.52) (S3,20.06) +- (0,1.65) (S4,156.84) +- (0,58.70)};
\legend{Rule-Based, DQN, AIF}
\end{axis}
\end{tikzpicture}
\caption{Mean per-step idle time per scenario (100 SUMO seeds, error bars $\pm 1$ std). Rule-based is worst throughout; AIF wins the full-disturbance Scenario~4 ($156.8$ vs $197.3$/$198.5$\,s for DQN/RB) and Scenario~2 ($20.2$ vs $23.1$\,s for DQN), while DQN leads in Scenarios~1 and~3, with much higher across-seed variance only in Scenario~1.}
\label{fig:scenario_idle}
\end{figure}
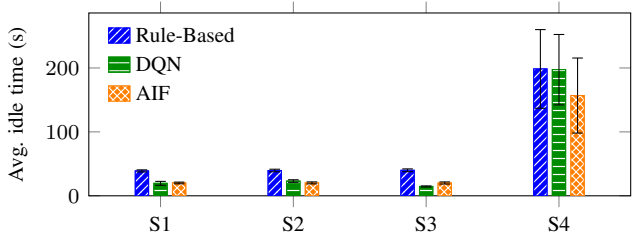

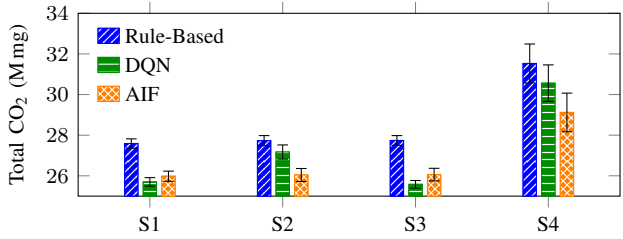
\begin{figure}[!tb]
\centering
\begin{tikzpicture}
\begin{axis}[scenbars, width=1\columnwidth, ylabel={Total CO\textsubscript{2} (M\,mg)}, ymin=25, ymax=34]
\addplot+[error bars/.cd, y dir=both, y explicit,
          error bar style={line width=0.4pt, black}]
    coordinates {(S1,27.59) +- (0,0.23) (S2,27.73) +- (0,0.25) (S3,27.74) +- (0,0.24) (S4,31.53) +- (0,0.96)};
\addplot+[error bars/.cd, y dir=both, y explicit,
          error bar style={line width=0.4pt, black}]
    coordinates {(S1,25.70) +- (0,0.21) (S2,27.17) +- (0,0.35) (S3,25.58) +- (0,0.19) (S4,30.56) +- (0,0.90)};
\addplot+[error bars/.cd, y dir=both, y explicit,
          error bar style={line width=0.4pt, black}]
    coordinates {(S1,25.98) +- (0,0.25) (S2,26.04) +- (0,0.32) (S3,26.06) +- (0,0.31) (S4,29.12) +- (0,0.95)};
\legend{Rule-Based, DQN, AIF}
\end{axis}
\end{tikzpicture}
\caption{Total CO\textsubscript{2} per scenario (100 SUMO seeds, error bars $\pm 1$ std). AIF is lowest in Scenarios~2 and~4 and within $0.5$\,M\,mg of DQN's lows in Scenarios~1 and~3; rule-based is always highest.}
\label{fig:scenario_co2}
\end{figure}

\begin{table*}[!htb]
\centering
\caption{Per-run cumulative metrics across 100 SUMO seeds: mean\,$\pm$\,std and CV per (scenario, controller). Idle in NS+EW waiting-seconds; CO\textsubscript{2} in M\,mg; Bus = priority service rate (\%); Sw.\ = phase switches.}
\label{tab:results_full}
\footnotesize
\setlength{\tabcolsep}{4pt}
\begin{tabular}{@{}llrcrcrcrc@{}}
\hline
\textbf{Sc.} & \textbf{Controller} &
\multicolumn{2}{c}{\textbf{Idle (s)}} &
\multicolumn{2}{c}{\textbf{CO\textsubscript{2} (M\,mg)}} &
\multicolumn{2}{c}{\textbf{Bus (\%)}} &
\multicolumn{2}{c}{\textbf{Sw.}} \\
 & & mean\,$\pm$\,std & CV & mean\,$\pm$\,std & CV & mean\,$\pm$\,std & CV & mean\,$\pm$\,std & CV \\
\hline
\multirow{3}{*}{1}
 & Rule-Based       & 14172\,$\pm$\,601   & 0.04 & 27.59\,$\pm$\,0.23 & 0.01 & 88.91\,$\pm$\,0.86 & 0.01 & 134\,$\pm$\,2 & 0.02 \\
 & DQN              &  7132\,$\pm$\,1328  & 0.19 & 25.70\,$\pm$\,0.21 & 0.01 & 84.80\,$\pm$\,1.11 & 0.01 & 220\,$\pm$\,4 & 0.02 \\
 & AIF &  7300\,$\pm$\,477   & 0.07 & 25.98\,$\pm$\,0.25 & 0.01 & 80.64\,$\pm$\,1.48 & 0.02 & 261\,$\pm$\,7 & 0.02 \\
\hline
\multirow{3}{*}{2}
 & Rule-Based       & 14400\,$\pm$\,735   & 0.05 & 27.73\,$\pm$\,0.25 & 0.01 & 88.22\,$\pm$\,1.21 & 0.01 & 134\,$\pm$\,3 & 0.02 \\
 & DQN              &  8385\,$\pm$\,740   & 0.09 & 27.17\,$\pm$\,0.35 & 0.01 & 84.59\,$\pm$\,1.12 & 0.01 & 262\,$\pm$\,5 & 0.02 \\
 & AIF &  7341\,$\pm$\,554   & 0.08 & 26.04\,$\pm$\,0.32 & 0.01 & 80.60\,$\pm$\,1.48 & 0.02 & 268\,$\pm$\,6 & 0.02 \\
\hline
\multirow{3}{*}{3}
 & Rule-Based       & 14520\,$\pm$\,796   & 0.05 & 27.74\,$\pm$\,0.24 & 0.01 & 88.19\,$\pm$\,1.11 & 0.01 & 134\,$\pm$\,3 & 0.02 \\
 & DQN              &  5305\,$\pm$\,318   & 0.06 & 25.58\,$\pm$\,0.19 & 0.01 & 85.63\,$\pm$\,1.30 & 0.02 & 266\,$\pm$\,6 & 0.02 \\
 & AIF &  7287\,$\pm$\,599   & 0.08 & 26.06\,$\pm$\,0.31 & 0.01 & 80.51\,$\pm$\,1.43 & 0.02 & 270\,$\pm$\,6 & 0.02 \\
\hline
\multirow{3}{*}{4}
 & Rule-Based       & 72247\,$\pm$\,22398 & 0.31 & 31.53\,$\pm$\,0.96 & 0.03 & 86.93\,$\pm$\,1.26 & 0.01 & 126\,$\pm$\,3 & 0.02 \\
 & DQN              & 71741\,$\pm$\,20030 & 0.28 & 30.56\,$\pm$\,0.90 & 0.03 & 89.65\,$\pm$\,1.09 & 0.01 & 177\,$\pm$\,6 & 0.03 \\
 & AIF & 56977\,$\pm$\,21318 & 0.37 & 29.12\,$\pm$\,0.95 & 0.03 & 82.02\,$\pm$\,1.33 & 0.02 & 278\,$\pm$\,7 & 0.03 \\
\hline
\end{tabular}
\end{table*}

The rule-based baseline performed worst in every scenario for idle time and CO\textsubscript{2}, and the active-inference vs DQN comparison flips with disturbance level. Under full-disturbance Scenario~4, AIF is clearly the best on both metrics ($56{,}977$\,s cumulative idle and $29.12$\,M\,mg CO\textsubscript{2}, vs $71{,}741$\,s and $30.56$\,M\,mg for DQN and $72{,}247$\,s and $31.53$\,M\,mg for rule-based). In the milder Scenarios~1--3, DQN attains the lowest mean cumulative idle time in Scenarios~1 and~3 (Scenario~1 marginally: $7{,}132$\,s vs AIF's $7{,}300$\,s; Scenario~3 decisively: $5{,}305$\,s vs $7{,}287$\,s), while AIF is the best in Scenario~2 ($7{,}341$\,s vs $8{,}385$\,s). However, the DQN headline in Scenario~1 hides a much larger seed-to-seed spread: idle standard deviation $1{,}328$\,s is $2.8\times$ AIF's $477$\,s and exceeds even the rule-based baseline's $601$\,s. AIF switches phase more frequently ($261$--$278$ per simulation) than DQN ($177$--$266$) or rule-based ($126$--$134$), with a lower bus priority service rate ($80.5$--$82.0$\% vs DQN's $84.6$--$89.7$\% and rule-based's $86.9$--$88.9$\%).

Across-seed idle-time dispersion splits two-and-two: AIF has the lowest std in Scenarios~1 and~2 ($477$\,s and $554$\,s), while DQN takes Scenarios~3 and~4 ($318$\,s and $20{,}030$\,s vs AIF's $599$\,s and $21{,}318$\,s). DQN's tightest distribution coincides with its lowest mean in Scenario~3 ($5{,}305$\,s). AIF's per-scenario mean idle is essentially scenario-invariant across Scenarios~1--3 ($7{,}287$--$7{,}341$\,s), so the ordering shift in Scenarios~1 and~3 ($168$\,s and $1{,}982$\,s gap to DQN) is driven by DQN improving rather than AIF degrading. Under Scenario~4 the idle-time CV climbs to $0.28$--$0.37$ and the distributions overlap substantially, yet AIF's mean of $56{,}977$\,s sits $14{,}764$\,s below DQN's $71{,}741$\,s, while DQN and rule-based's $72{,}247$\,s differ by only $506$\,s. CO\textsubscript{2}, bus priority, and switch counts remain low-dispersion (CV $\leq 0.03$) throughout.

\begin{figure}[!htb]
\centering
\begin{tikzpicture}
\begin{axis}[
    width=1\columnwidth, height=4.6cm,
    xlabel={Simulation time (s)}, ylabel={Cumulative idle (s)},
    xmin=0, xmax=3630, ymin=0, ymax=100000,
    label style={font=\footnotesize},
    tick label style={font=\footnotesize},
    legend cell align=left,
    legend style={font=\footnotesize, draw=none, fill=none, at={(0.02,0.98)}, anchor=north west},
    every axis plot/.append style={thick},
    grid=major, grid style={dotted, gray!30},
]
\addplot[blue!60, dashed, line width=0.4pt, forget plot] table[x=t, y=rb_lo] {cumidle_s4.dat};
\addplot[blue!60, dashed, line width=0.4pt, forget plot] table[x=t, y=rb_hi] {cumidle_s4.dat};
\addplot[green!55!black!60, dashed, line width=0.4pt, forget plot] table[x=t, y=dqn_lo] {cumidle_s4.dat};
\addplot[green!55!black!60, dashed, line width=0.4pt, forget plot] table[x=t, y=dqn_hi] {cumidle_s4.dat};
\addplot[orange!70, dashed, line width=0.4pt, forget plot] table[x=t, y=ai_lo] {cumidle_s4.dat};
\addplot[orange!70, dashed, line width=0.4pt, forget plot] table[x=t, y=ai_hi] {cumidle_s4.dat};
\addplot[blue]           table[x=t, y=rb]  {cumidle_s4.dat}; \addlegendentry{Rule-Based}
\addplot[green!55!black] table[x=t, y=dqn] {cumidle_s4.dat}; \addlegendentry{DQN}
\addplot[orange]         table[x=t, y=ai]  {cumidle_s4.dat}; \addlegendentry{AIF}
\draw[red!60, dashed, thick] (axis cs:2000,0) -- (axis cs:2000,100000);
\draw[red!60, dashed, thick] (axis cs:2300,0) -- (axis cs:2300,100000);
\end{axis}
\end{tikzpicture}
\caption{Cumulative idle time under Scenario~4 across 100 SUMO seeds (mean line, $\pm 1$ std band). Red dashed lines mark the accident ($t = 2000$--$2300$~s). After the accident, the rule-based baseline and DQN accumulate idle at nearly the same rate, while AIF grows visibly more slowly and ends $14{,}764$\,s and $15{,}270$\,s below DQN and rule-based, respectively.}
\label{fig:cumulative_idle_s4}
\end{figure}
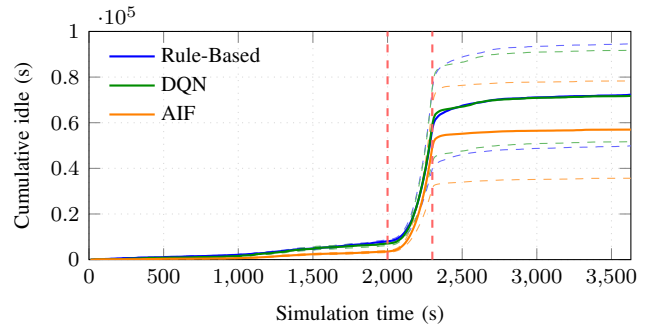

The Scenario~4 cumulative-idle traces (Fig.~\ref{fig:cumulative_idle_s4}) show AIF accumulating idle more slowly than DQN and rule-based from the start ($639$\,s vs $1{,}568$\,s vs $2{,}114$\,s at $t = 1000$\,s; AIF at $30$\% of rule-based), with the gap widening sharply over the injected accident at $t = 2000$\,s and the subsequent high-demand period: rule-based and DQN finish on top of each other ($72{,}247$\,s and $71{,}741$\,s) while AIF ends clearly below ($56{,}977$\,s). The per-step idle spike at the accident peaks at $t = 2{,}300$~s with similar mean heights ($4{,}139$\,s, $4{,}242$\,s, $4{,}633$\,s for AIF, rule-based, DQN), so the cumulative gap comes from slower idle accumulation throughout rather than from differences in spike height.

\begin{figure}[!htb]
\centering
\begin{tikzpicture}
\begin{axis}[
    width=1\columnwidth, height=4.6cm,
    xlabel={Simulation time (s)}, ylabel={Cumulative switches},
    xmin=0, xmax=3630, ymin=0,
    label style={font=\footnotesize},
    tick label style={font=\footnotesize},
    legend cell align=left,
    legend style={font=\scriptsize, draw=none, fill=none, at={(0.02,0.98)}, anchor=north west, row sep=-2pt},
    every axis plot/.append style={thick},
    grid=major, grid style={dotted, gray!30},
]
\addplot[blue!60, dotted, line width=0.4pt, forget plot] table[x=t, y=rb_lo] {switches_s4.dat};
\addplot[blue!60, dotted, line width=0.4pt, forget plot] table[x=t, y=rb_hi] {switches_s4.dat};
\addplot[green!55!black!60, dotted, line width=0.4pt, forget plot] table[x=t, y=dqn_lo] {switches_s4.dat};
\addplot[green!55!black!60, dotted, line width=0.4pt, forget plot] table[x=t, y=dqn_hi] {switches_s4.dat};
\addplot[orange!70, dotted, line width=0.4pt, forget plot] table[x=t, y=ai_lo] {switches_s4.dat};
\addplot[orange!70, dotted, line width=0.4pt, forget plot] table[x=t, y=ai_hi] {switches_s4.dat};
\addplot[blue]                   table[x=t, y=rb]  {switches_s4.dat}; \addlegendentry{Rule-Based, S4}
\addplot[green!55!black]         table[x=t, y=dqn] {switches_s4.dat}; \addlegendentry{DQN, S4}
\addplot[orange]                 table[x=t, y=ai]  {switches_s4.dat}; \addlegendentry{Active Inf., S4}
\addplot[blue, dashed]           table[x=t, y=rb]  {switches_s2.dat}; \addlegendentry{Rule-Based, S2}
\addplot[green!55!black, dashed] table[x=t, y=dqn] {switches_s2.dat}; \addlegendentry{DQN, S2}
\addplot[orange, dashed]         table[x=t, y=ai]  {switches_s2.dat}; \addlegendentry{Active Inf., S2}
\end{axis}
\end{tikzpicture}
\caption{Cumulative phase-switch counts under Scenario~2 (dashed) and Scenario~4 (solid, $\pm 1$ std across 100 seeds). AIF and DQN are on par in Scenario~2 ($268$ vs $262$), while AIF clearly leads in Scenario~4 ($278$ vs $177$); rule-based stays lowest in both ($134$, $126$). AIF and rule-based switch at near-identical rates across scenarios while DQN switches less under full disturbance, indicating elevated AIF switching is structural to greedy EFE minimization rather than  a response to the disturbance.}
\label{fig:switches_s2_s4}
\end{figure}
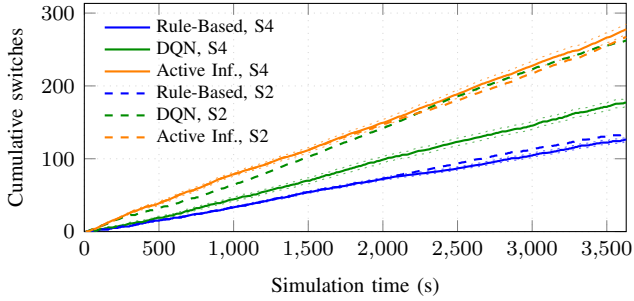

The cumulative phase-switch counts (Fig.~\ref{fig:switches_s2_s4}) show AIF and DQN essentially on par in Scenario~2 ($268$ vs $262$), while AIF clearly leads in Scenario~4 ($278$ vs $177$); rule-based stays lowest throughout ($134$ in Scenario~2, $126$ in Scenario~4). The separations are established within the first few hundred seconds and grow monotonically rather than emerging around the disturbance window, evidence that AIF's elevated switching under full disturbance is policy-level rather than disturbance-driven.

\section{Discussion} \label{sec:discussion}
Our results suggest that AIF can achieve competitive performance in simulated traffic control when appropriately parameterized, trading wins with a pretrained DQN under stable conditions (DQN ahead on mean cumulative idle in Scenarios~1 and~3, AIF ahead in Scenario~2) while substantially outperforming a rule-based heuristic throughout (roughly halving its cumulative idle in Scenarios~1--3), and outperforming both decisively in the full-disturbance Scenario~4. This parallels \cite{lanillos2021active}, who survey AIF controllers for robotic manipulation and report that combining state estimation and control within a single free-energy objective matches or outperforms task-specific baselines. As a point of reference in this domain, \cite{gao2017adaptive} report that a DQN-based controller reduces vehicle delay by up to 47\% relative to a longest-queue-first policy and 86\% relative to fixed-time control; our results place AIF within a similar performance range without requiring an explicitly specified reward function or gradient-based optimization.

The asymmetry in pretraining matters. The DQN baseline trains a per-scenario policy over $500$--$1{,}000$ episodes, so its mean idle time decreases in Scenarios~1 ($7{,}132$\,s vs AIF's $7{,}300$\,s) and~3 ($5{,}305$\,s vs $7{,}287$\,s) reflect scenario-specific fitting rather than zero-shot generalization. AIF has no pretraining stage: it acts at every step through its specified generative model, with the EFE objective driving uncertainty-reducing exploration online via its epistemic-value. That a training-free controller still matches DQN in Scenario~1 ($168$\,s gap), beats it in Scenario~2 ($7{,}341$\,s vs $8{,}385$\,s), and outperforms it decisively in Scenario~4 ($56{,}977$\,s vs $71{,}741$\,s) is the main empirical observation in Table~\ref{tab:results_full}, consistent with challenges in \cite{chen2022real} that limited sensing and nonstationary demand drive sim-to-real degradation in RL-based controllers. Across-seed idle-time dispersion splits two-and-two between the two controllers (AIF tightest in Scenarios~1 and~2, DQN tightest in Scenarios~3 and~4), so the model-based formulation absorbs sensing and demand noise comparably to a per-scenario-trained value-function approximator rather than uniformly better.

Unlike DQN, every stage of the AIF pipeline (posterior belief, per-action prediction, pragmatic and epistemic EFE components, action posterior) is inspectable at each decision step, addressing the verification difficulty that \cite{liao2025symlight} identify for deep reinforcement learning-based traffic controllers.

AIF carries the lowest bus priority service rate in every scenario ($80.5$--$82.0$\% vs DQN's $84.6$--$89.7$\% and rule-based's $86.9$--$88.9$\%). The pragmatic value rewards low congestion across the whole approach, and because buses are a small fraction of total vehicles, the EFE-optimal action typically serves the higher car load even when a bus is queued opposite, rather than weighting bus presence as a separate objective.

\subsection{Limitations and Future Work}

The
simulated conditions designed to reflect realistic traffic scenarios, providing a controlled yet representative setting for comparison. Extending the framework to multi-intersection networks and hardware-in-the-loop deployment are natural next steps. The baseline set covers the most established approaches in this domain. The current parameter configuration was selected to ensure fair comparison, though a comprehensive systematic exploration of alternative settings may demonstrate further performance insights. Future work could broaden comparisons to policy-gradient algorithms and exploit AIF's built-in exploration to investigate uncertainty 
reduction across algorithms beyond task reward.

\section{Conclusion} \label{sec:conclusion}

We presented an AIF controller for a signalized intersection under noisy, nonstationary IoT sensing conditions and evaluated it against a rule-based heuristic and a DQN across four scenarios of increasing disturbance. AIF achieved the lowest idle time and CO\textsubscript{2} emissions under full disturbance without scenario-specific training, while remaining competitive in milder conditions. The results suggest that the epistemic value term in the EFE objective provides a structural advantage under partial observability and nonstationarity, making AIF a viable alternative to reinforcement learning for adaptive traffic control. Future work may extend the framework to multi-intersection networks, broaden baseline comparisons, and progress toward hardware-in-the-loop deployment.

\bibliographystyle{IEEEtran}
\bibliography{biblio}

\end{document}